\documentclass[twoside,a4paper]{article}
\usepackage{fleqn,espcrc2}
\usepackage{graphicx}
\usepackage{latexsym}
\usepackage{euscript}
\newcommand{\ifig}[1]{\includegraphics[height=75mm,width=80mm]{#1}}
\newcommand{\bc}{\begin{center}}
\newcommand{\ec}{\end{center}}
\newcommand{\be}{\begin{equation}}
\newcommand{\ee}{\end{equation}}
\newcommand{\bsp}{\begin{split}}
\newcommand{\esp}{\end{split}}
\newcommand{\bea}{\begin{eqnarray}}
\newcommand{\eea}{\end{eqnarray}}
\newcommand{\ba}{\begin{eqnarray}}
\newcommand{\ea}{\end{eqnarray}}
\newcommand{\bas}{\begin{eqnarray*}}
\newcommand{\eas}{\end{eqnarray*}}

\newcommand{\simge}{\ \lower-
1.2pt\vbox{\hbox{\rlap{$>$}\lower5pt
\vbox{\hbox{$\sim$}}}}\ }

\newcommand{\AC} {{\cal{A}}}
\newcommand{\AmS}{{\protect\the\textfont2
  A\kern-.1667em\lower.5ex\hbox{M}\kern-.125emS}}
\title{ 
 Results on the
 Gluon Propagator in  Lattice Covariant Gauges
}
\author{L. Giusti\address{Boston University - Department of Physics,
590 Commonwealth Avenue, Boston MA 02215 USA},
 M. L. Paciello\address{INFN, Sezione di Roma 1,
 P.le A. Moro 2, I-00185 Roma, Italy},
 S. Petrarca\thanks{Speaker at the Conference}$^{\rm b~}$\address{Dipartimento di Fisica, Universit\`a di Roma ``La
                     Sapienza''}%
        ,
        C. Rebbi$^{\rm a}$,
B. Taglienti$^{\rm b}$
                     }

\begin{document}

\begin{abstract}
We illustrate preliminary results on the gluon propagator
computed in  generic covariant lattice gauges in the  quenched approximation
with the Wilson action. We have applied a recently proposed procedure to fix
a generic covariant gauge on the lattice.
We make a comparison among results obtained at two different values
of the gauge parameter and in the standard Landau's gauge.
\vspace{1pc}
\end{abstract}

\maketitle

We  present preliminary data on the gluon propagator in generalized 
covariant gauges.  The motivation for considering a generalized
gauge-fixing procedure stems, just as in the continuum, from 
the necessity of studying the gauge 
dependence of  the Green's functions.
In the covariant gauges it is possible to change the gauge fixing
varying the value of the gauge parameter. 
In the continuum, the generic covariant gauges satisfy the 
following condition:
\be
{\partial_{\mu}}{A_{\mu}}^{G_\lambda}  (x)=\Lambda_\lambda (x) 
\label{eq:dinamic1}
\ee
where $\Lambda_{\lambda}(x)$  are  matrices belonging to the Lie algebra of the
SU(3) group and $ \lambda$ is the gauge parameter  which controls the width 
of the gaussian distribution of the $\Lambda$'s. 

The corresponding gauge fixing procedure on the lattice is implicitly 
defined by the following expression for the expectation value of a gauge
dependent operator ${\cal O}$:
\bea
\langle{\cal O}\rangle&=&\frac{1}{Z}\int d\Lambda \int dU  \nonumber \\
& &{\cal O}({U^{G_\lambda}})
e^{-\frac{1}{\lambda}\sum_x
Tr(\Lambda^2)}
e^{-\beta S(U)} 
\label{eq:omedio}
\eea
where $S(U)$ is the Wilson lattice gauge invariant action 
and $G_{\lambda }$ is the gauge transformation that enforces
the gauge condition.

After having generated a set of $\Lambda$ according to a Gaussian distribution,
the gauge fixing procedure follows very closely the widely used 
Landau's gauge fixing: the gauge transformation $G(x)$ which 
rotate the links $U_\mu(x)$ into the gauge-fixed configuration 
is obtained by the numerical minimization of a functional 
$F[G]$, chosen in such a way that its minimum corresponds to 
a gauge configuration satisfying the gauge condition.
The functional form we adopt in the continuum reads:
\be
F[G]=\int d^4x\mbox{\rm Tr}\left[(\partial_{\mu}A^{G_\lambda}_{\mu}-\Lambda)^2  
\right]\; .  
\label{eq:cov11}
\ee
The discretization of this expression requires a particular care (driven
discretization) in order to obtain a functional form suitable to be
minimized by the standard gauge-fixing algorithms.
More details on this covariant gauge-fixing method can be found in 
\cite{giusti}, \cite{covar}.

The stationary points of the above functional correspond to the following
gauge condition in the continuum:
\be
D_{\nu}\partial_{\nu}(\partial_{\mu}A^{G_\lambda}_{\mu}-\Lambda)=0
\label{eq:casino}
\ee
It should be noted that, while for $\Lambda=0$ eq.~(\ref{eq:casino})
reduces to the Landau gauge condition (apart from the presence
of possible spurious solutions, i.e.~solutions to 
$\partial_{\mu}A^{G_\lambda}_{\mu}-\Lambda=\rho$ with 
$D_{\nu}\partial_{\nu} \rho=0$ which, however, we did not encounter
in our study), the same is not true of the lattice construction,
where the equations that follow from the minimization of the
functional with $\Lambda=0$ differ from the lattice
Landau gauge condition by terms $O(a)$.  Thus the comparison
of the results we will obtain for $\Lambda=0$ with those
obtained in the Landau gauge will be a meaningful check on our procedure.

We gauge-fixed 221 SU(3) configurations with $V=16^3\times32$ and $\beta=6$
taking the gauge parameter $\lambda=0, 8$ where $\lambda=0$ 
has been considered in order to make a comparison with the Landau gauge.
The gauge configurations were retrieved from the repository 
at the ``Gauge Connection'' (http://qcd.nersc.gov/).
The calculation has been performed on Boston University's  
Origin2000 and took approximately 19000 CPU hours.
We imposed a gauge fixing quality factor  $\theta< 10^{-6}$.
With these gauge-fixed configurations we computed the correlator 
of gluon propagator
\be
D_{\mu\nu}(x-y)=<A_\mu(x) A_\nu(y)>
\label{eq:defprop}
\ee
where we adopted for $A$ the standard expression 
\be \label{eq:standard}
A_{\mu} (x) 
\equiv  \left[{{U_{\mu} (x) - U_{\mu}^{\dagger} (x)}\over
{2 i a g_0}}\right]_{Traceless}\; .
\ee 
Discussions on the effect of different lattice definitions of the gluon field
in the context of numerical calculations can be found in 
refs.~\cite{giusto} and~\cite{dubnap}. The general properties of 
lattice operators were studied in ref.~\cite{testa}.

\begin{figure}[h]
\bc
\ifig{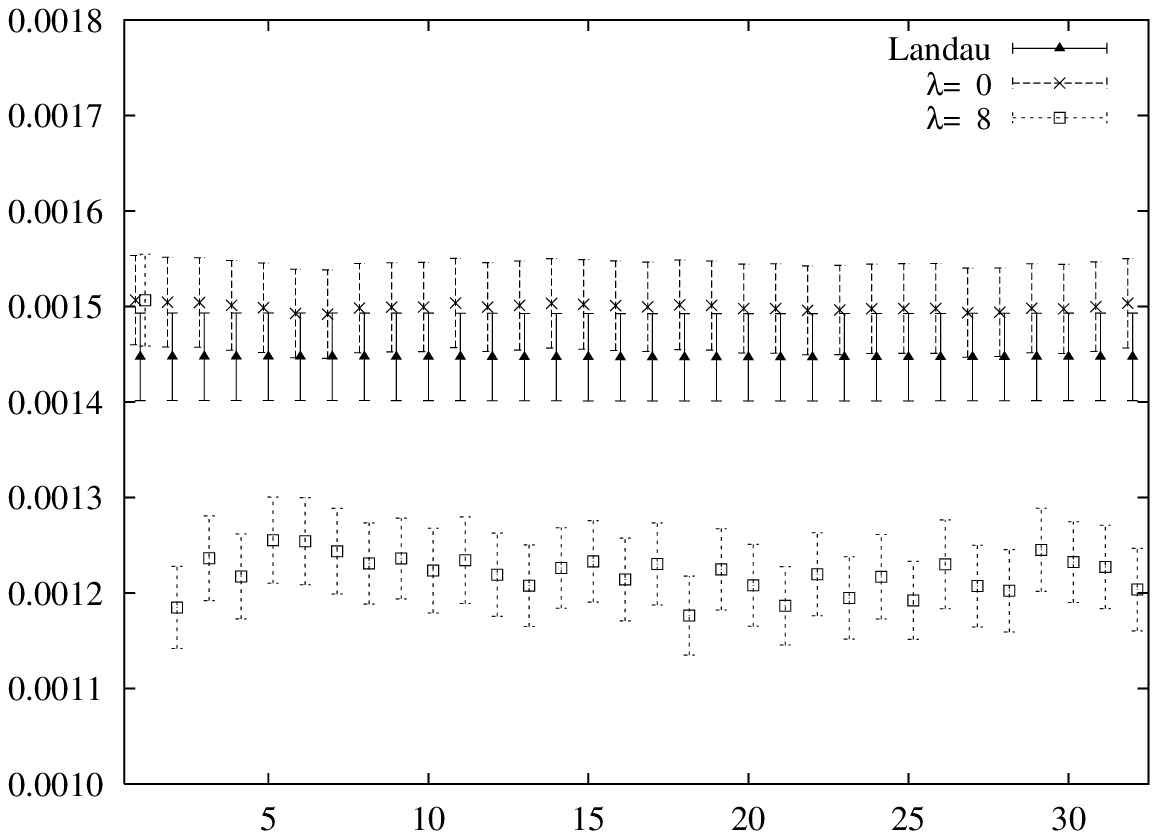}
\caption{\small{Behaviour of 
  $\langle\AC_0\AC_0\rangle(t)$ 
 as a function of time for 
a set of 221 thermalized $SU(3)$  configurations at $\beta=6.0$ with a 
volume $V\times T=16^3\times 32$. The data have been
slightly displaced in $t$ to help the eye, the errors have
 been obtained with the jacknife method.}}
\label{fig:a0}
\ec
\end{figure}

As usual in gluon propagator 
calculations~\cite{manrep}  we study the quantity 
\be
\langle \AC_0\AC_0\rangle (t) \equiv \frac{1}{V^2}  
 \sum_{{\bf x},{\bf y}} Tr \langle  A_0({\bf x},t)A_0({\bf y},0)\rangle \; .
\label{eq:A0A0}\\
\ee
which is displayed in Fig.~\ref{fig:a0}.
This correlator should be a constant in the case of Landau's
gauge-fixing and its behaviour is indicative of the quality
of the gauge fixing procedure. 
Figure~\ref{fig:a0} shows that even in the case $\lambda=8$
the above correlator is almost a costant.  The case $\lambda=0$ is 
equivalent to the Landau's one with a  different value of the constant. 
 
\begin{figure}[h]
\bc
\ifig{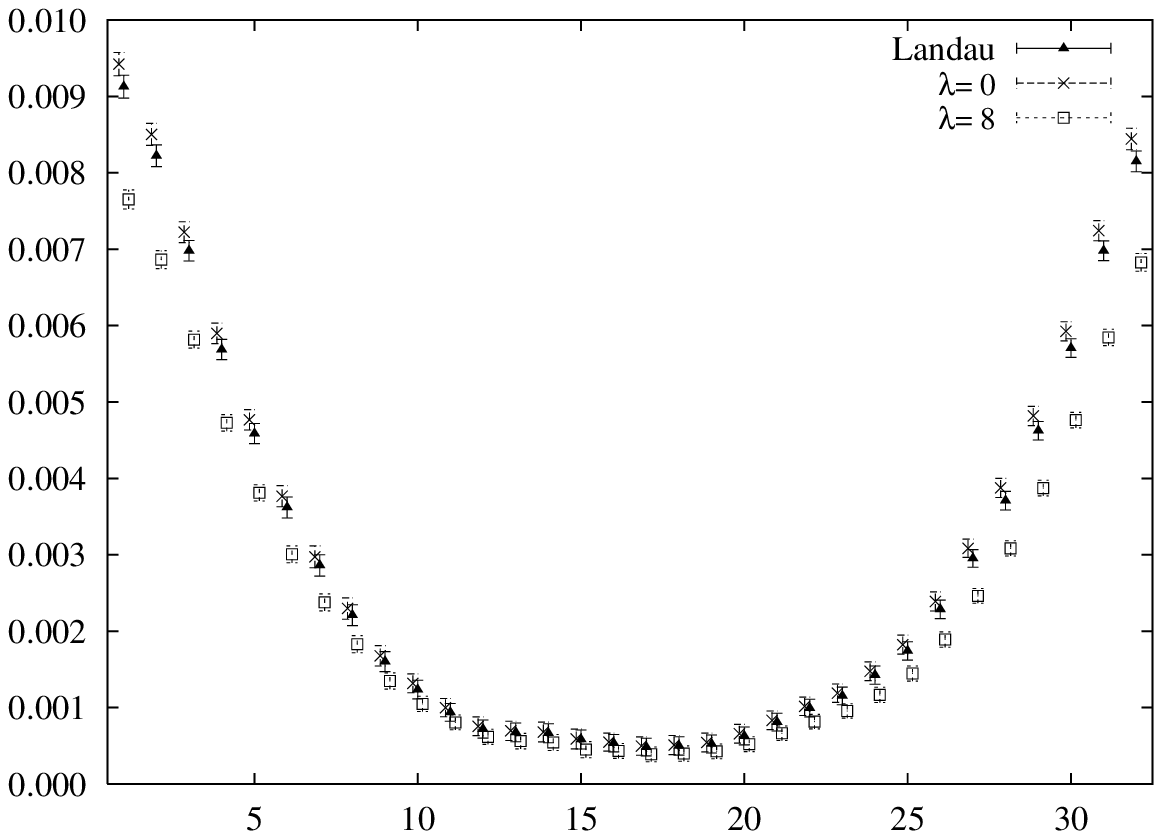}
\caption{\small{Gluon propagator eq.~(\ref{eq:AiAi}) as a function of~t 
 for a simulation with 221 
 thermalized configurations at $\beta=6$ and $V=16^3 \times 32$. The
 apparent asymmetry with respect to the midpoint of the lattice
 is only due to the slight shift of two sets of points along the time 
 axis.}}
%propagatore in x
\label{fig:ai}
\ec
\end{figure}

In Fig.~\ref{fig:ai} we exhibit the behaviour of the propagator 
\be
\langle \AC_i\AC_i\rangle (t) \equiv \frac{1}{3 V^2}  
 \sum_{i}\sum_{{\bf x},{\bf y}} Tr \langle  A_i({\bf x},t)A_i({\bf y},0)\rangle
\label{eq:AiAi}\\
\ee
as a function of $t$.
%propagatore in x 

The results of this simulation show clearly 
that the gauge dependence of the gluon propagator (eq.~(\ref{eq:AiAi})) 
on $\lambda$ cannot be simply understood as a rescaling of the data,
as one might have argued of the data on a smaller lattice,
reported in the very preliminar analysis of ref.~\cite{dubna}. 
We also observe that the data for the Landau gauge fixing
and for $\lambda=0$ are compatible within errors.
We proceed by showing the gluon propagator in Fourier space.
The transformed field $A_\mu(k)$ is given by
\be
A_{\mu}(k)=e^{-{i \over 2}k_\mu}\sum_x e^{-i (k \cdot x)} A_\mu(x)
\label{eq:propk}\\
\ee
where $k_\mu={2 \pi \over L_\mu} n_\mu$ with $L_\mu=16$ or $32$ 
for spatial and temporal indices respectively, $n_\mu$ are integers.
Following suggestions from
lattice perturbation theory~\cite{skulle}, in order
to get rid of a well-known lattice artifact we express all the quantities
as a function of the variable $q$ 
$$q_\mu=2 \sin({k_\mu \over 2}) $$
rather than $k$.  This kinematic correction is needed to reduce the 
anisotropy of the data when displayed as function of $k$.
The Fourier transform of the gluon propagator in eq.~(\ref{eq:defprop}) is:
\be
D_{\mu\nu}(q)={1 \over V}< A_\mu(q) A_\nu{^{\dagger}}(q) > 
\label{eq:propq}\\
\ee
We analyzed the data in terms of the conventional distinction of 
transverse and longitudinal parts $D_T$ and $D_L$:
\be
D_{\mu\nu}(q)=(\delta_{\mu\nu} - {q_\mu q_\nu \over q^2})D_T(q) 
+{q_\mu q_\nu \over q^2}{D_L(q) \over q^2}
\label{eq:lontra}\\
\ee
%terza figura
\begin{figure}[h]
\bc
\ifig{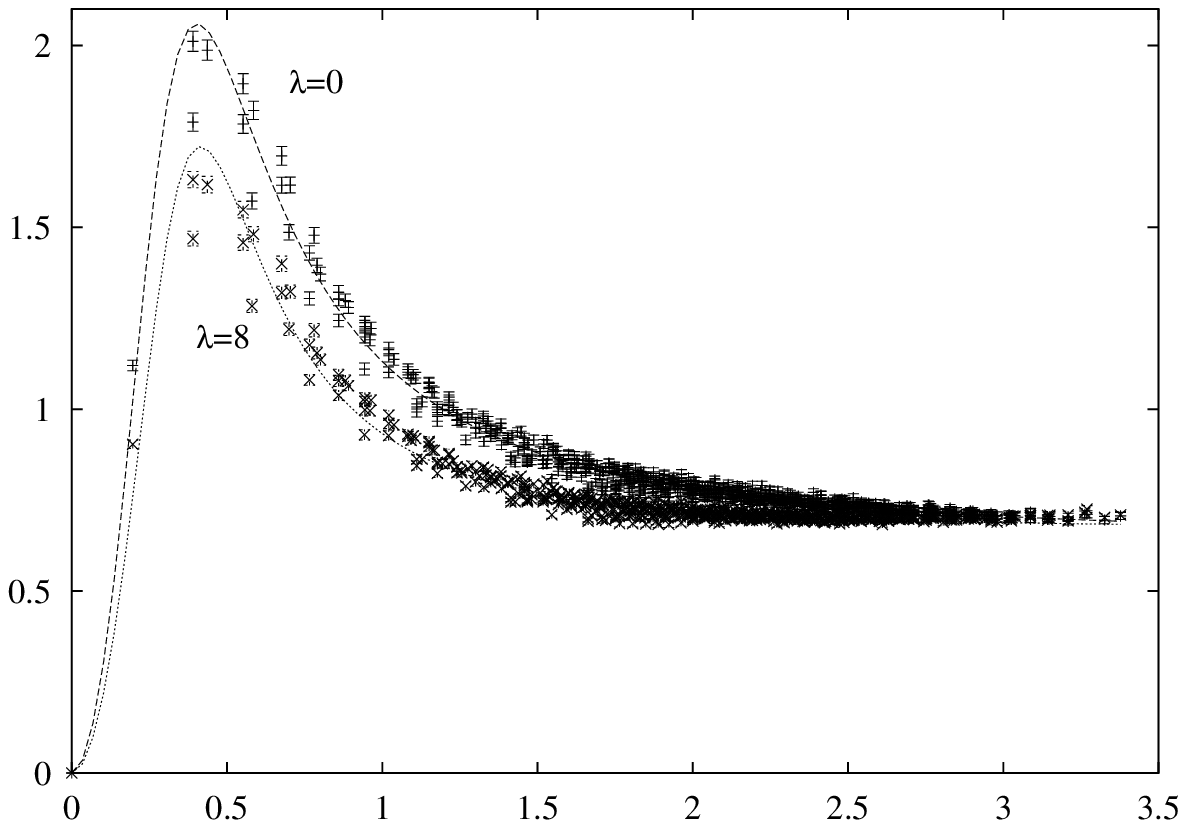}
\caption{\small{ Transverse part of the gluon propagator $q^2 D_T(q)$ in
covariant gauges as a function of $q$.
The two sets of data refer to $\lambda=0$ and $\lambda=8$,
221 thermalized $SU(3)$  configurations at $\beta=6.0$ with a 
volume $V\times T=16^3\times 32$.}}
\label{fig:tran}
\ec
\end{figure} 
%propagatore longitudinale quarta
\begin{figure}[h]
\bc
\ifig{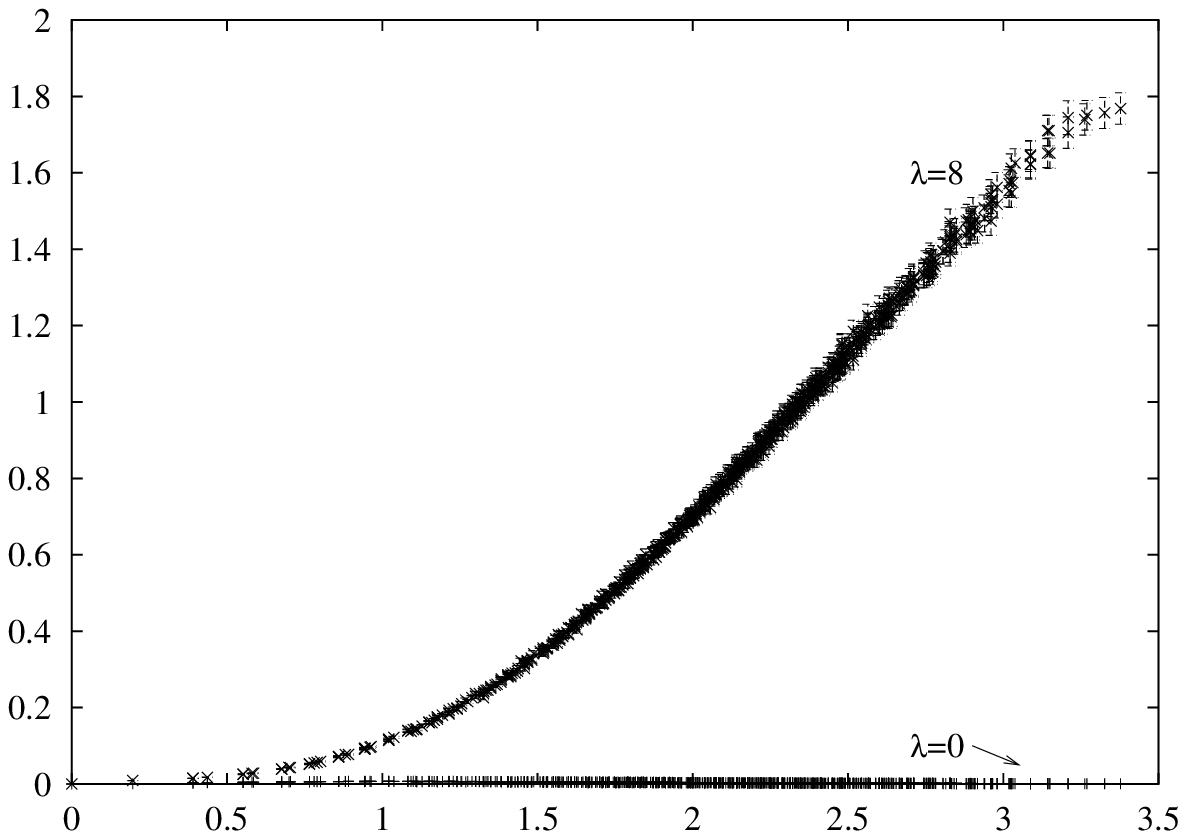}
\caption{\small{Longitudinal part of the gluon propagator as a function of $q$ at 
$\lambda=0$ and $\lambda=8.$ 
}}
\label{fig:longitud}
\ec
\end{figure}
In Fig.~\ref{fig:tran} we present the behaviour of the transverse
part (times $q^2$) of the gluon propagator in covariant gauges, for the two
values of the gauge parameter $\lambda=0$, equivalent to
the Landau's gauge, and $\lambda=8$, as a function of
$q=\sqrt{ q^2}$. The data have been averaged over
the $Z_3$ group and only values corresponding to 
$k_0={2 \pi \over 32} (0 \div 12)$ and
$k_i={2 \pi \over 16} (0 \div 6)$ are reported.
We see that at large $q^2$ the data at the two values of the gauge parameter
tend to coincide, while the gluon propagator at low
$q^2$ exhibits a clear gauge dependence. 
The data with $\lambda=8$ show a lower  peak
than those with $\lambda=0$, but the peak position stays fixed.
In Fig.~\ref{fig:longitud} we illustrate the behaviour 
of the longitudinal part $D_L$ as a function of $q$.

The data at $\lambda=0$ vanish as one expects for Landau's gauge. At 
$\lambda=8$ the situation is different and a signal different to zero
is observed with a clear increase for increasing $q$. This differs from
the predictions of pertubation theory, where the longitudinal part is 
a constant proportional to the value of the gauge
parameter.

In summary, our gauge fixing procedure can be implemented for medium
to large volume simulation; the gauge fixing obtained with our algorithm 
behaves as one expects in the case of $\lambda=0$,
reproducing the features of Landau's gauge. 
The data of the gluon propagator
are new and show a clear gauge dependendence; 
the transverse and longitudinal part 
will be analyzed in more details in a future publication.
 
We warmly thank Massimo Testa for many fruitful discussions.
We thank the Center for Computational Science of Boston 
University where the largest part of this computation was done.
L.~G.~and C.~R.~have been supported in part under DOE grant
DE-FG02-91ER40676.

\end{document}
%===========================================================================
%==========================================================================
%===========================================================================